\documentclass[prb,twocolumn,amsmath,amssymb,superscriptaddress,showpacs]{revtex4}
\usepackage{graphicx}
\usepackage{dcolumn}
\usepackage{bm}

\usepackage{enumerate}

\begin{document}

\title{Magnetization of Quantum Dots: A Measure of Anisotropy and 
the Rashba Interaction}
\author{Siranush Avetisyan}
\affiliation{Department of Physics and Astronomy,
University of Manitoba, Winnipeg, Canada R3T 2N2}
\author{Tapash Chakraborty$^\ddag$}
\affiliation{Department of Physics and Astronomy,
University of Manitoba, Winnipeg, Canada R3T 2N2}
\author{Pekka Pietil\"ainen}
\affiliation{Department of Physics/Theoretical Physics,
University of Oulu, Oulu FIN-90014, Finland}

\begin{abstract}
The magnetization of anisotropic quantum dots in the presence of the Rashba spin-orbit interaction has
been studied for three interacting electrons in the dot. We observe unique behaviors of magnetization
that are direct reflections of the anisotropy and the spin-orbit interaction parameters independently or 
concurrently. In particular, there are saw-tooth structures in the magnetic field dependence of the 
magnetization, as caused by the electron-electron interaction, that are strongly modified in the presence 
of large anisotropy and high strength of the spin-orbit interactions. We report the temperature dependence
of magnetization that indicates the temperature beyond which these structures due to the interactions 
disappear. Additionally, we found the emergence of a weak sawtooth structure in magnetization in the high 
anisotropy and large spin-orbit interaction limit that was explained as a result of merging of two low-energy 
curves when the level spacings evolve with increasing values of the anisotropy and the spin-orbit
interaction strength.
\end{abstract}
\maketitle

The magnetization of quantum confined planar electron systems, e.g. quantum dots (QDs), or
artificial atoms \cite{maksym,qdbook,heitmann} is an important probe that reflects entirely 
on the properties of the energy spectra. This is a thermodynamical quantity that for the QDs 
has received some experimental attention \cite{wilde_10,schwarz,delft_98}, particularly after 
the theoretical prediction that the electron-electron interaction is directly manifested in 
this quantity \cite{QD_mag}. In addition to the large number of theoretical studies reported 
in the literature on the electronic properties of isotropic quantum dots, there has been lately 
some studies on the anisotropic quantum dots, both theoretically \cite{madhav,others} and 
experimentally \cite{elliptic_expt}. Theoretical studies of the magnetization of elliptical 
QDs have also been reported \cite{QD_mag_aniso}. Effects of the Rashba spin-orbit interaction 
(SOI) \cite{rashba} on the electronic properties of isotropic \cite{rashba_tc} and anisotropic 
quantum dots \cite{aniso_rashba} have been investigated earlier. An external electric field 
can induce the Rashba spin-orbit interaction which couples the different spin states and
introduces level repulsions in the energy spectrum \cite{rashba_tc,aniso_rashba,hong-yi}. This 
coupling is an important ingradient for the burgeoning field of semiconductor spintronics, in
particular, for quantum computers with spin degrees of freedom as quantum bits 
\cite{grundler,science}. Three-electron quantum dots are particularly relevant in this context
\cite{quant_comput,aron}. Here we report on the magnetic field dependence of the magnetization of 
an anisotropic QD containing three interacting electrons in the presence of the Rashba SOI. Our 
present work clearly indicates how the magnetization of the QDs uniquely reflects the influence 
of anisotropy and the Rashba SOI, both concurrently as well as individually as the strengths of 
the SOI and the anisotropy are varied independently. The temperature dependence of magnetization 
is also studied here, where we noticed the gradual disappearence of the interaction induced 
structures in magnetization with increasing temperature. Another important feature that we found 
in our present study is the emergence of a weak sawtooth structure in magnetization in the high 
anisotropy and large spin-orbit interaction limit that we explain as a result of merging of two 
low-energy curves when the level spacings evolve with increasing parameters. With the help of the
theoretical insights presented here, experimental studies of magnetization will therefore provide 
valuable information on the inter-electron effects, the Rashba spin-orbit coupling and the degree 
of anisotropy of the quantum dots.

At zero temperature the magnetization $\cal M$ of the QD is defined as ${\cal M} = -\frac{\partial 
E^{}_{\rm g}}{\partial B}$ where $E^{}_{\rm g}$ is the ground state energy of the system 
\cite{QD_mag,magneti}. We have studied the magnetic field dependence of $\cal M$ by evaluating 
the expectation value of the magnetization operator ${\hat m} = -\frac{\partial {\cal H}}{\partial B}$,
where $\cal H$ is the system Hamiltonian. Since the Columb interaction is independent of $B$, 
$\hat m$ would be just a one-body operator, i.e., we can ignore the interaction part from 
the Hamiltonian. The Hamiltonian of a single-electron system subjected to an external magnetic 
field with the vector potential ${\bf A}=\frac12 B(-y,x)$, the confinement potential, and the Rashba 
SOI is
\begin{eqnarray*}
{\cal H} &=& \frac1{2m^{}_e}\left({\bf p} - \frac ec {\bf A}\right)^2
+ \tfrac12 m^{}_e\left(\omega_x^2x^2 + \omega_y^2y^2\right) \\
&+& \frac{\alpha}{\hbar}\left[{\bm \sigma}\times({\bf p} - \frac ec{\bf A})\right]^{}_z 
+ \tfrac12 g\mu^{}_{B}B\sigma^{}_z.\\
\end{eqnarray*}
The first term of the Hamiltonian is the kinetic energy, which can be written as
$$K = \frac{1}{2m^{}_e}\left(p_x^2 + p_y^2 + \frac{eB}{c}(yp^{}_x - xp^{}_y)
+ \frac{e^{2}B^2}{4c^2}(y^2 + x^2)\right).$$
The SOI part (third term) is
$$H^{}_{\rm SO}
=\frac{\alpha}{\hbar}\left[\sigma^{}_x(p^{}_y-\frac{eB}{2c}x) - \sigma^{}_y(p^{}_x 
+ \frac{eB}{2c}y)\right],$$
while the second and the last term correspond to the confinement potential and the Zeeman
term, respectively. We then need to evaluate the expectation value of the magnetization operator
\begin{eqnarray*}
\hat m &=& -\frac{\partial H}{\partial B} \\
= &-&\frac1{2m^{}_e}\frac{e}{c}\left((yp^{}_x-xp^{}_y) + \frac{eB}{2c}(y^2+x^2)\right) \\
&+& \frac{e\alpha}{2c\hbar}(\sigma^{}_{x}x+\sigma^{}_{y}y)-\tfrac12 g\mu^{}_B\sigma^{}_z,\\
\end{eqnarray*}
with respect to the interacting electron states. We should, however, point out that the 
energy spectra in the present studies were evaluated for the Hamiltonian with the Coulomb 
interaction $V^{}_c=e^2/\epsilon r$ included, as in our earlier work \cite{aniso_rashba_int} 
but now for three interacting electrons. Here $\epsilon$ is the background dielectric constant.

\begin{center}
\begin{figure}
\includegraphics[width=8.2cm]{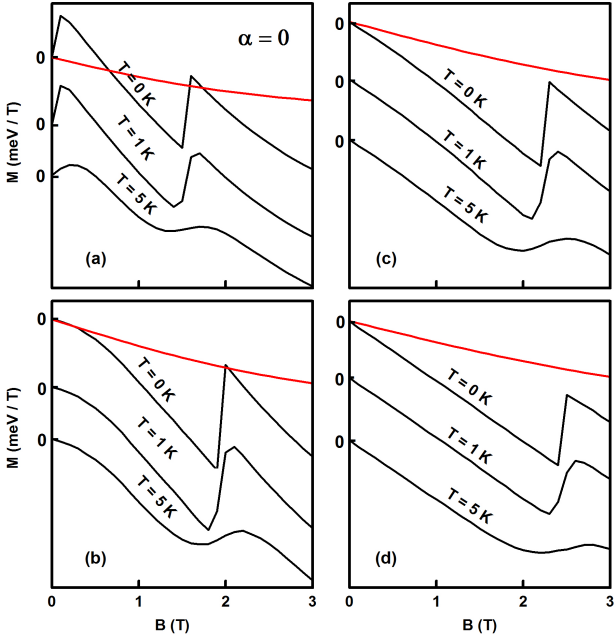}
\caption{\label{fig:alpha_mag=0} Temperature dependence of magnetization of a three-electron
anisotropic dot without the Rashba SOI $(\alpha=0)$. The results are for $\omega^{}_x=4$ meV
and (a) $\omega^{}_y=4.1$ meV, (b) $\omega^{}_y=6$ meV, (c) $\omega^{}_y=8$ meV,
and (d) $\omega^{}_y=10$ meV.  The zero-temperature magnetization curve for the
non-interacting system is also shown in red.
}
\end{figure}
\end{center}

We have also studied the finite-temperature behavior of the magnetization, following the 
thermodynamical model discussed earlier \cite{thermo}. Since we are investigating the system
with a fixed number of electrons, we use the canonical ensemble. The temperature dependence 
of the magnetization was therefore evaluated from the thermodynamic expression
\begin{equation}
{\cal M}=\sum^{}_m\frac{\partial E^{}_m}{\partial B}\,{\rm e}^{-E^{}_m/kT}/\sum^{}_m
{\rm e}^{-E^{}_m/kT},
\end{equation}
where the partial derivatives were evaluated, as explained above, as the expectation values of
the magnetization operator in the interacting states labelled by $m$. 
In elliptical confinements, the mutual Coulomb interaction is handled by the numerical scheme
elucidated previously \cite{aniso_rashba_int}, i.e., we diagonalize the many-body Hamiltonian
in the basis consisting of non-interacting many-body states, which are constructed by the SO coupled
single-particle spinors. These spinors are in turn, as the result of the diagonalization of the SO 
Hamiltonian, expressed as superpositions of the fundamental 2D oscillator spinors 
$$|n^{}_x,n^{}_y;s^{}_z\rangle=|n^{}_x,n^{}_y\rangle|s^{}_z\rangle.$$
Here $n^{}_x$ and $n^{}_y$ are the oscillator quantum numbers and $|s^{}_z\rangle$ stands for the 
spinors 
\begin{eqnarray*}
|+1\rangle = |\uparrow\rangle = \left(\begin{array}{c}
                        1 \\
                        0
                      \end{array}\right)
\end{eqnarray*}
\begin{eqnarray*}
|-1\rangle = |\downarrow\rangle = \left(\begin{array}{c}
                        0 \\
                        1
                      \end{array}\right).
\end{eqnarray*}
We see that in the end the magnetization evaluation reduces to a many-fold summation of the matrix 
elements
\begin{eqnarray*}
&&\langle n'_x,n'_y;s'_z|\hat m|n^{}_x,n^{}_y;s^{}_z\rangle \\
& =& -\frac{e}{2m^{}_ec}\langle n_x^{'}, n_y^{'};s^{}_z|yp^{}_x-xp^{}_y|n^{}_x, n^{}_y
;s^{}_z\rangle  \\
&-& \frac{e^2B}{4m^{}_e c^2}<n_x^{'}, n_y^{'};s^{}_z|y^2+x^2|n^{}_x, n^{}_y;s^{}_z\rangle            \\
& + & \frac{\alpha e}{2c\hbar}<n_x^{'}, n_y^{'};s_z^{'}|\sigma^{}_xx
+ \sigma^{}_yy|n^{}_x, n^{}_y;s^{}_z\rangle \\
&-& \tfrac1{2}g\mu^{}_B\langle n_x^{'}, n_y^{'};s_z^{'}|\sigma^{}_z|n^{}_x, n^{}_y;s^{}_z\rangle,
\end{eqnarray*}
which is now susceptible to direct numerical evaluation.

\begin{center}
\begin{figure}
\includegraphics[width=8.2cm]{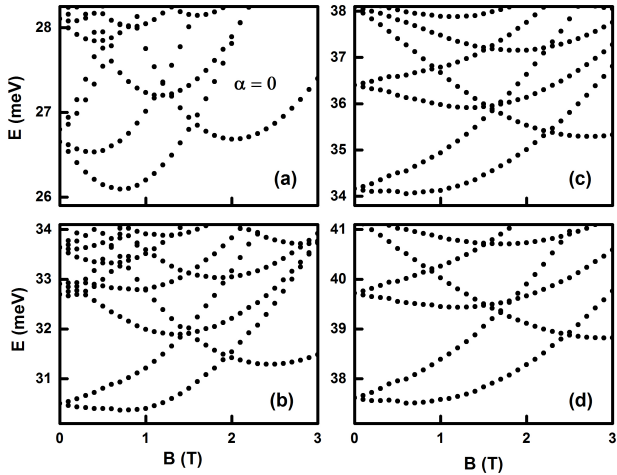}
\caption{\label{fig:alpha_en=0} Energy levels of a three-electron anisotropic dot
without the Rashba SOI $(\alpha=0)$. The results are for $\omega^{}_x=4$ meV and (a)
$\omega^{}_y=4.1$ meV, (b) $\omega^{}_y=6$ meV, (c) $\omega^{}_y=8$ meV,
and (d) $\omega^{}_y=10$ meV.
}
\end{figure}
\end{center}

In our numerical investigations, we chose the InAs quantum dot that shows strong
Rashba effects \cite{rashba_tc,aniso_rashba,aniso_rashba_int}. In this case, the
relevant parameters are: $\epsilon=15.15, g=-14, m^{}_e=0.042.$ The energy spectra are 
shown in Fig. 2, Fig. 4, and Fig. 6 for various values of the SO coupling strength 
$\alpha$ and for different values of the anisotropy. For $B=0$ the ground states
are two-fold degenerate no matter how strong the SO coupling is or how anisotropic
the QD becomes. Interestingly, contrary to our expectations, at non-zero magnetic fields 
most of the level crossings of the energy spectra do not turn into anticrossings when the
SO coupling is turned on. Only for a strong value of the Rashba parameter $\alpha$ 
($\alpha=40$) [Fig. 6 (a), (b)] the level crossings transform to level repulsions.
However, when the QD becomes more anisotropic those level repulsions reappear as level
crossings [Fig. 6 (c),(d)]. 

\begin{center}
\begin{figure}
\includegraphics[width=8.2cm]{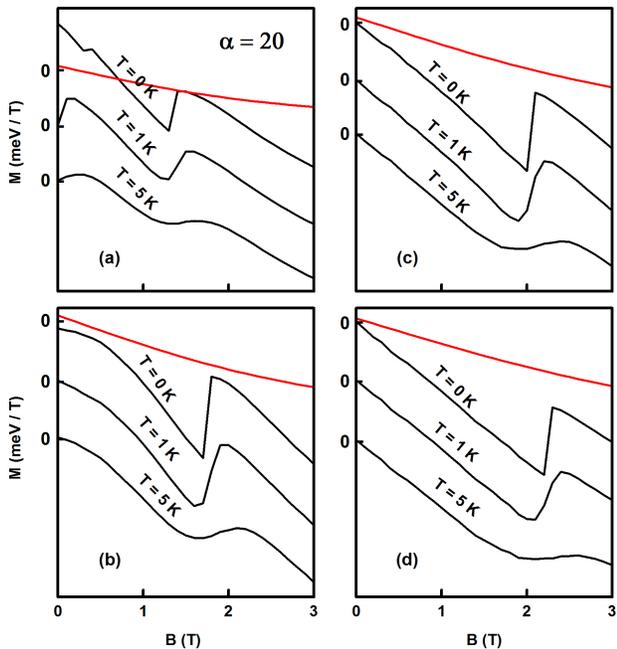}
\caption{\label{fig:alpha_mag=20} Same as in Fig.~\ref{fig:alpha_mag=0}, but for
$\alpha=20$ meV.nm.}
\end{figure}
\end{center}

Our results for the magnetic field dependence of the magnetization for anisotropic QDs are
shown in Fig. 1, Fig. 3 and Fig. 5, calculated both with and without (red curves) the
Coulomb interaction between the electrons for various values of the SO coupling strength
and for various values of anisotropy. A major difference between the the non-interacting system 
and the interacting system can be found in the magnetization results: while there is no
structure present in the non-interacting cases (red curves), there are prominent structures
for the interacting systems. As it was predicted in earlier theoretical works \cite{QD_mag}
(and confirmed in our present work), the electron-electron interaction causes saw-tooth 
structure in the magnetic field dependence of the magnetization, which is a consequence 
of the change in the ground state energy from one magic angular momentum to another (in the
case of isotropic QDs) \cite{qdbook}. An interesting behavior of magnetization that should be 
pointed out here is that with increasing strength of the Rashba SO parameter $\alpha$ the jump 
in magnetization at the level crossings in the energy spectra, moves to lower magnetic fields, 
while increasing anisotropy of the QD pushes the jump in magnetization to higher magnetic fields. 
For InAs elliptical QDs this shift is at most $\sim 1$ Tesla, when $\alpha$ is increased from zero 
to 40 meV.nm and $\omega^{}_y$ is varied from 4.1 - 10 meV. Therefore, low-field magnetization 
measurements of the QDs could be a direct probe of the SO coupling strength.

\begin{center}
\begin{figure}
\includegraphics[width=8.2cm]{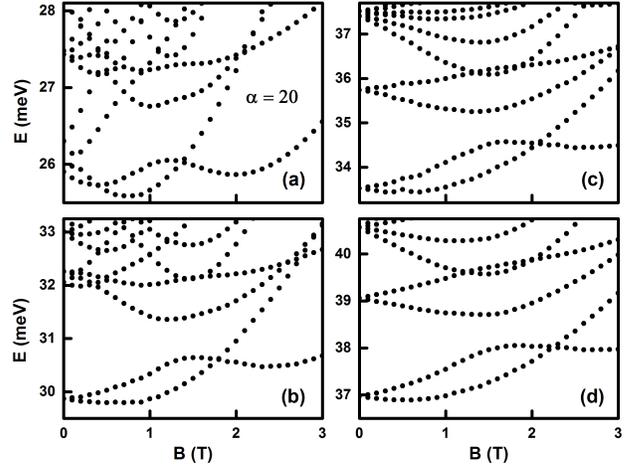}
\caption{\label{fig:alpha=20} Same as in Fig.~\ref{fig:alpha_en=0}, but for
$\alpha=20$ meV.nm.}
\end{figure}
\end{center}

The main feature of the temperature dependence of magnetization is that, as the temperature
is increased the saw-tooth structure of the magnetization curve gradually disappears (Fig. 1
and Fig. 3). An important point to notice here is that the jump in magnetization slowly
vanishes even in the absence of increasing temperature for an anisotropic QD [Fig. 5 (a), (b)]
which is a result of large SO coupling. However, an emergent small jump in magnetization
is again visible (at $T=0$ K) for strong anisotropic QDs and large SOI [Fig. 5 (c), (d)] which is 
clearly the result of two low-lying energy levels crossing near 1.2 Tesla [Fig. 6 (c), (d)]
(marked by circles).

\begin{center}
\begin{figure}
\includegraphics[width=8.2cm]{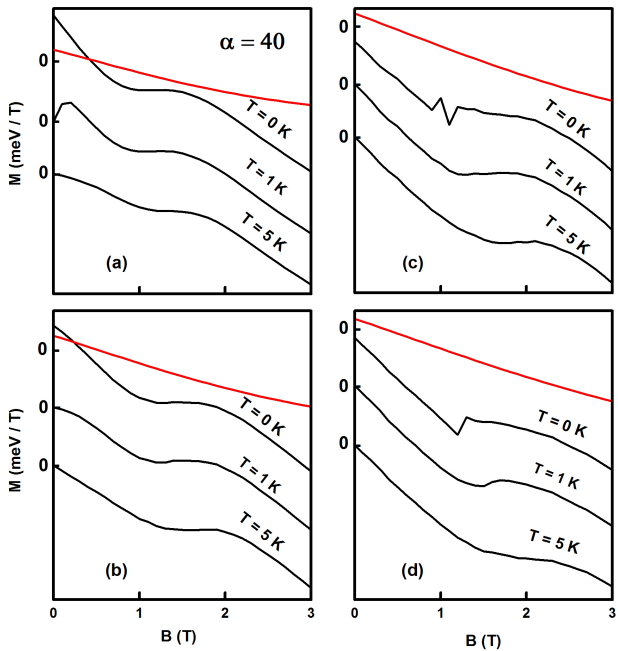}
\caption{\label{fig:alpha_mag=40} Same as in Fig.~\ref{fig:alpha_mag=0}, but for
$\alpha=40$ meV.nm.}
\end{figure}
\end{center}

In the absence of the SO coupling the magnetization is invariant under the time 
reversal, which implies that the derivative of energy with respect to the magnetic 
field must vanish at $B=0$. This means a vanishing magnetization at $B=0$. However, 
non-zero SO coupling breaks the time-reversal symmetry and in that case the derivative of 
energy with respect to the magnetic field can be discontinuous at $B=0$, which would imply 
non-vanishing magnetization at $B=0$ [Fig. 4 and Fig. 6].

\begin{center}
\begin{figure}
\includegraphics[width=8.2cm]{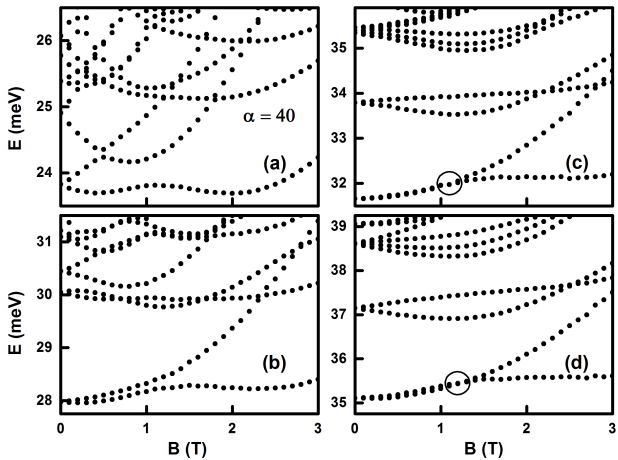}
\caption{\label{fig:alpha=40} Same as in Fig.~\ref{fig:alpha_en=0}, but for
$\alpha=40$ meV.nm. The circles in (c) and (d) indicate the level crossings
that lead to new structures in Fig. 5 (c) and (d) at $T=0$ K.}
\end{figure}
\end{center}

To summarize: we have reported here detailed and accurate studies of the magnetization
of anisotropic quantum dots with interacting electrons in the presence of the Rashba SOI. 
The Coulomb interaction in the presence of the spin-orbit coupling exhibits a very strong 
effect on magnetization, particularly in the presence of strong anisotropy by introducing 
large saw-tooth structures in the magnetic field dependence of the magnetization, which
is weakened by increasing temperature. Interestingly, there is also the emergence of this 
structure in the high anisotropy and large SOI limit that is explained as due to merging of 
two low-energy curves when the level spacings evolve with increasing parameters. Any further 
extension of the present work for a larger system would be computationally very challenging, 
but we expect that the novel features observed in the present work will be displayed in 
that case. That would be the subject of our future publications. Armed with the theoretical 
insights presented here, an experimental probe of magnetization in anisotropic quantum dots 
will undoubtedly provide valuable information about the inter-electron strength, the 
strength of the QD anisotropy, as well as the SOI strength in the quantum dot.

The work was supported by the Canada Research Chairs Program of the Government of Canada.

\end{document}